\documentclass[aip,apl,reprint,a4paper]{revtex4-1}

\usepackage{graphicx}
\usepackage{textcomp}
\usepackage{verbatim}

\begin{document}


\title{Independence of surface morphology and reconstruction during the thermal preparation of perovskite oxide surfaces}



\author{Maren J\"ager}
\author{Ali Teker}
\author{Jochen Mannhart}
\author{Wolfgang Braun}
\email[corresponding author, ]{w.braun@fkf.mpg.de}
\affiliation{Max Planck Institute for Solid State Research, Heisenbergstr.\ 1, 70569 Stuttgart, Germany}


\date{\today}

\begin{abstract}
Using a CO$_2$ laser to directly heat the crystals from the back side, SrTiO$_3$ substrates may be thermally prepared in situ for epitaxy without the need for ex-situ etching and annealing.
We find that the formation of large terraces with straight steps at 900--1100\,$^\circ$C is independent from the formation of the ideal surface reconstruction for epitaxy, which requires temperatures in excess of 1200\,$^\circ$C to complete.
The process may be universal, at least for perovskite oxide surfaces, as it also works, at different temperatures, for LaAlO$_3$ and NdGaO$_3$, two other widely used oxide substrate materials.
\end{abstract}


\maketitle

Epitaxial oxide layers offer a wide range of functionalities, as they exhibit novel electronic properties when combined in heterostructures\cite{ogale2013functional,watanabe1995reproducible,mathews1997ferroelectric,ohtomo2004high,lee2005strong,cook2014double,ramesh2008whither,christen2008interfaces,schlom2008athin,mele2015oxide,boschker2017quantum}.
The quality of such epitaxial films crucially depends on the quality of the substrate they are grown on.
The choice of substrates, and in particular their surface preparation before deposition, therefore are decisive for the synthesis of high-quality heterostructures.

The relevance of substrate preparation has been recognized in 1994\citep{kawasaki1994atomic}.
The preparation procedure of standard substrates\cite{sanchez2014tailored,biswas2017atomically} such as SrTiO$_3$ typically involves etching in HF solution and an extended bakeout in oxygen at ambient pressure\cite{kawasaki1994atomic,koster1998quasi,kawasaki1994atomic,koster1998influence,liang1994structures}.
This process takes place ex situ, consists of several process steps and is therefore time consuming.
Furthermore, it involves hydrofluoric acid which requires strict safety procedures.

An in-situ substrate preparation process similar to the oxide desorption of compound semiconductors would be desirable as it is cleaner, faster, safer and better to control.
We have therefore examined the in-situ thermal preparation of SrTiO$_3$ and similar oxide substrates.
Substrate preparation using thermal annealing has been studied early on in the context of surface preparation for surface structure characterization\cite{liang1994structures, jiang1999c6x2}, with more recent dedicated surveys aimed at obtaining good templates for epitaxial growth\cite{gunnarsson2009evaluation,connell2012preparation}.
Here, we find that annealing the pre-cleaned substrate in an oxygen partial pressure in the single digit Pa range at high temperatures produces surfaces that are indistinguishable from surfaces prepared with the traditional procedure and may be used in-situ for the subsequent deposition of heterostructures.

The substrates are heated in a pulsed laser deposition (PLD) chamber using a CO$_2$ laser system which is also used during deposition.
The laser used\citep{trumpf} is capable of delivering a maximum power of 1.1\,kW at a wavelength of 9.27\,\textmu m.
The beam is shaped into a top-hat profile and then projected at variable size onto the sample plane using zoom optics.
It is reflected into the growth chamber by a Bragg mirror with a transmission window from 6.7 to 8.3\,\textmu m, allowing on-axis access of a pyrometer for temperature control.
The control pyrometer emissivity is set to 1.0, any stated temperature values therefore should be lower bounds to the actual temperatures.
A fast switching input allows for abrupt modulation of the laser beam at any power.
Details of the sample heater setup will be published separately.

\begin{figure}
\includegraphics[width=\columnwidth]{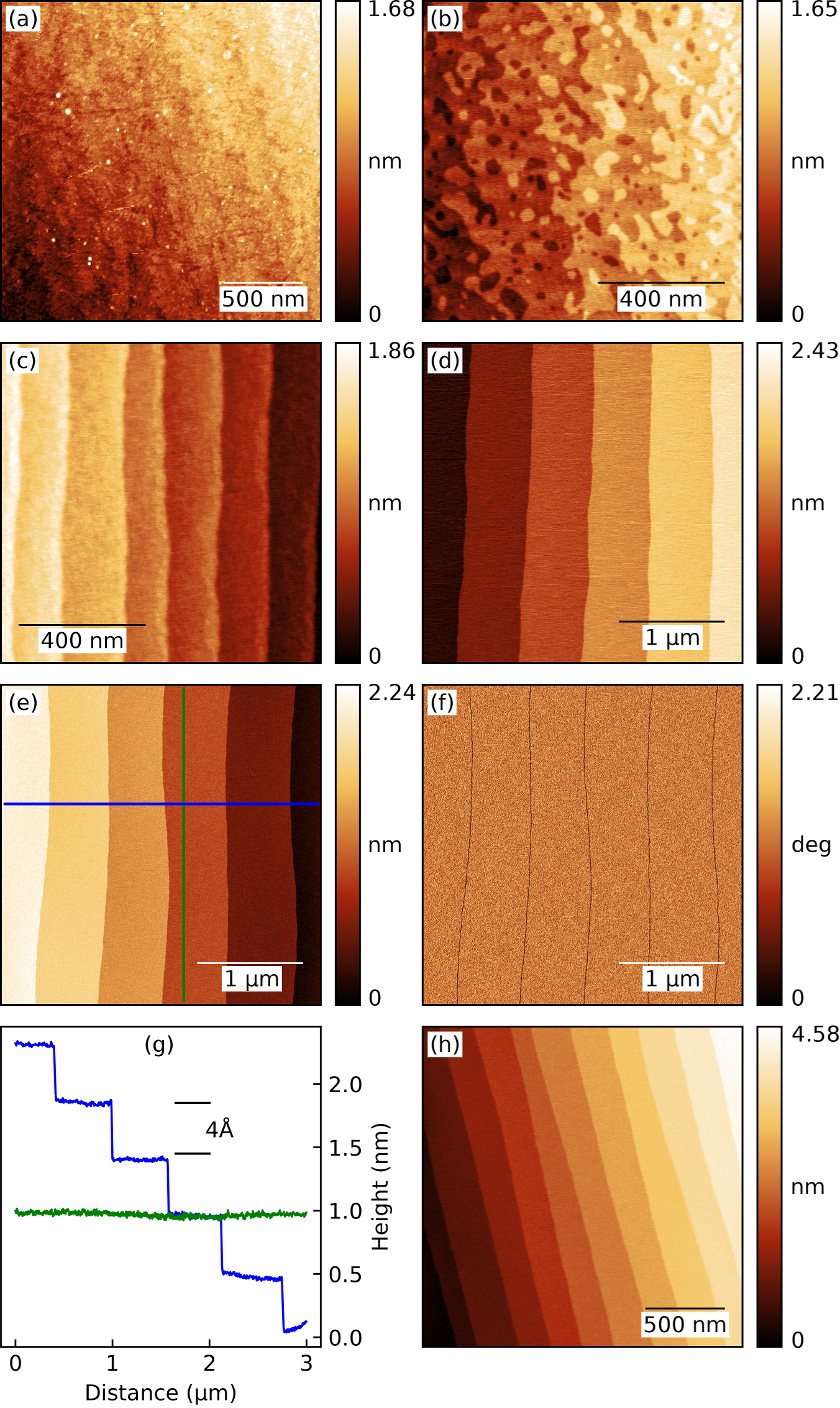}%
\caption{AFM images of typical SrTiO$_3$ (001) surfaces before (a) and during the thermal treatment at (b) 870\,$^\circ$C, (c) 1100\,$^\circ$C, (d) 1300\,$^\circ$C, and (e) 840\,$^\circ$C.
                Lines in (e) indicate the positions of the height profiles shown in (g).
                The phase image (f) reveals contrast at the terrace edges only, indicating a uniform reconstruction of the surface.
                Image (h) shows a surface slowly cooled to room temperature from 840\,$^\circ$C.
                All images taken at 300\,K, (b--f) after rapidly quenching the crystals from the specified temperatures.\label{fig:thermalprep}}%
\end{figure}
Sample preparation and growth has been performed in an advanced PLD system using an x-y-z motion controlled target table and a KrF excimer ablation laser\citep{coherent} with a pulse width of 25\,ns and a wavelength of 248\,nm.
The background gas consisted of purified molecular oxygen (nominal purity 6.0).
The target-sample distance was 56\,mm.
Samples were 5$\times$5$\times$0.5\,mm$^3$ and 5$\times$5$\times$0.3\,mm$^3$ in size, with a ground finish back side to reduce the reflectivity for the heating laser.

\begin{figure}
\includegraphics[width=\columnwidth]{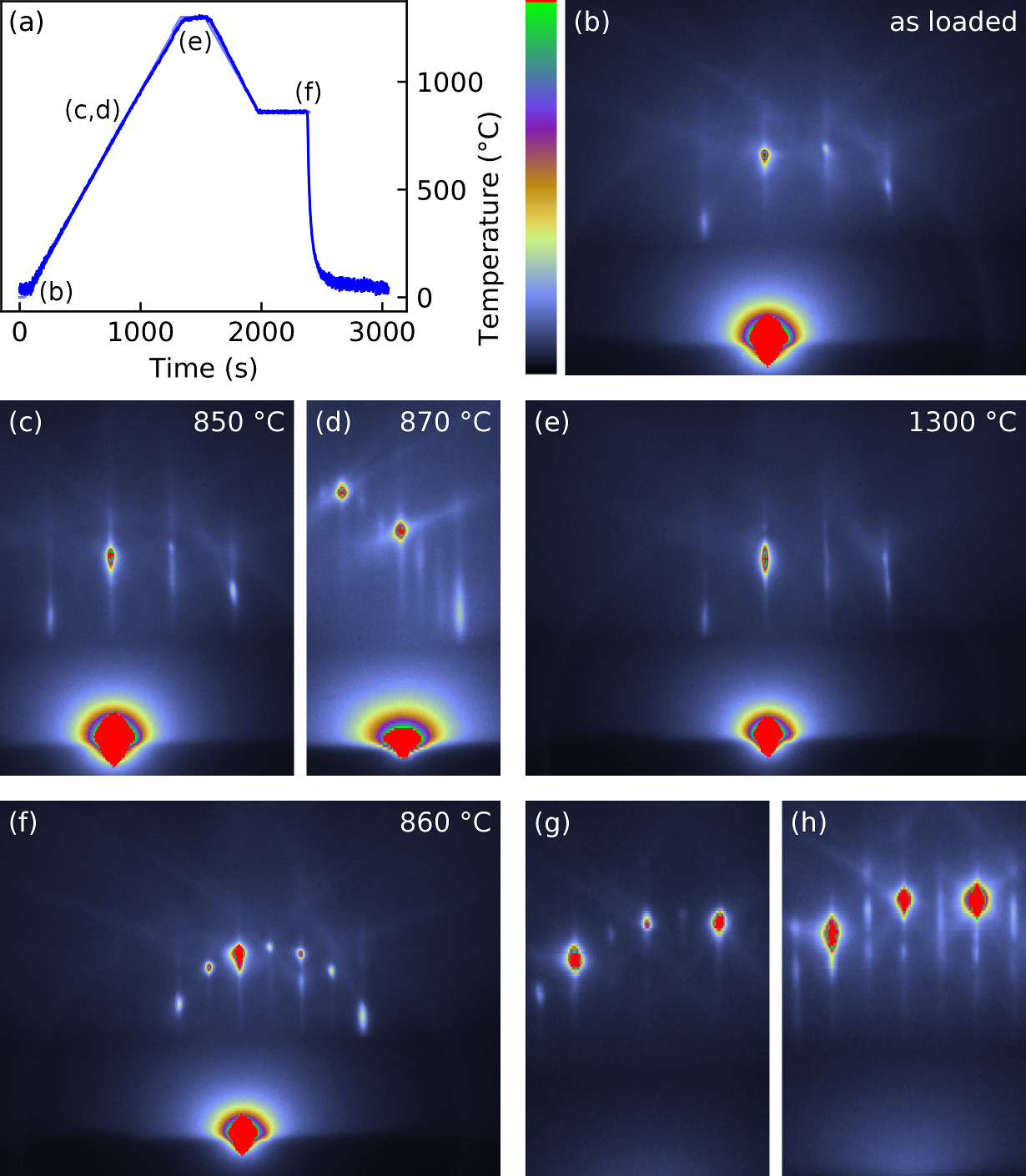}%
\caption{Measured temperature profile (a) and false-color RHEED intensity distributions of thermally prepared substrates at various stages during the process (b-f), and, for comparison, of a chemically prepared substrate (g,h).
The false-color conversion bar for (b) to (h) is shown next to (b).\label{fig:rheed}}%
\end{figure}

Figure~\ref{fig:thermalprep} shows atomic force microscopy (AFM) images\cite{asylum} taken ex-situ from SrTiO$_3$ (001) surfaces before, at various stages during, and after the preparation.
The substrates are first wiped with a synthetic fiber tissue soaked in propanol, then sonicated in acetone and propanol baths for 10\,min each.
They are then sonicated in ultrapure water for another 10\,min and blown dry with nitrogen.
Fig.~\ref{fig:thermalprep}(a) displays the morphology of the surface after these steps.
As indicated by the height scale, the surface is already quite flat in this state, however with significant lateral disorder and a large defect density.
The substrate is then introduced into the growth chamber and heated to 1300\,$^\circ$C at a rate of 1\,K/s, held at 1300\,$^\circ$C for 200\,s, and cooled to the growth temperature, here 840\,$^\circ$C, again with 1\,K/s.
The measured temperature profile of this process is given in Fig.~\ref{fig:rheed}(a).
The oxygen background pressure during the process is 7.5\,Pa (0.075\,mbar).

\begin{figure}
\includegraphics[width=\columnwidth]{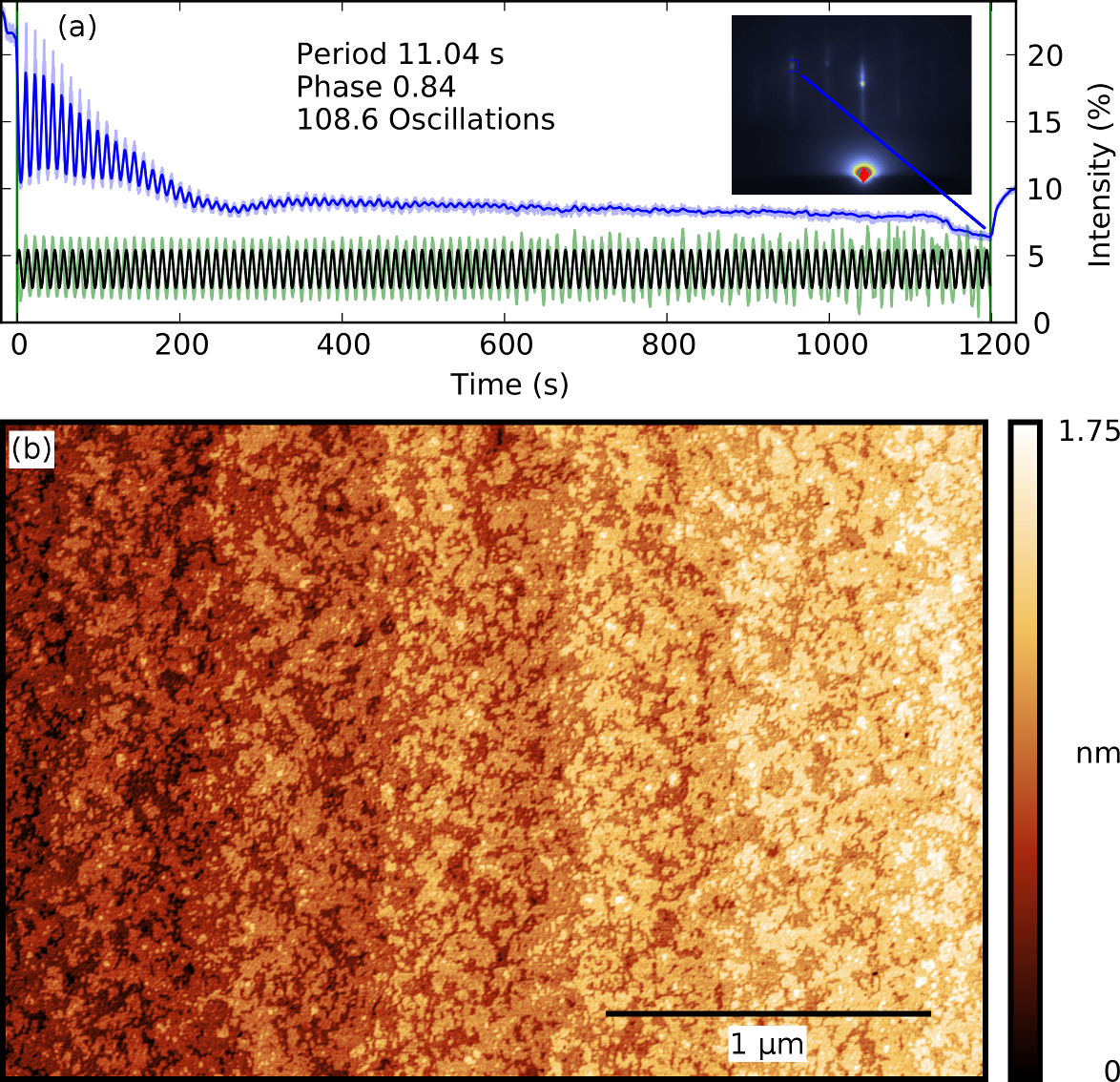}%
\caption{RHEED intensity oscillations (a) and AFM topography snapshot (b) of the final surface morphology of a SrTiO$_3$ (001) film grown homoepitaxially on a thermally prepared substrate.
                The insert shows the RHEED pattern after growth, with the area used for the oscillation measurement indicated by the blue frame.\label{fig:growth}}%
\end{figure}

At various stages during the process, the heating laser is turned off abruptly, and the substrate rapidly cools to room temperature, at which AFM images are taken\footnote{The temperature profile shown in Fig.~\ref{fig:rheed}(a) corresponds to the sequence used for panel Fig~\ref{fig:growth}(e--g)}.
At 870\,$^\circ$C during the heating phase (Fig.~\ref{fig:thermalprep}(b)), slightly above the typical homoepitaxial growth temperature, terraces and islands have already formed with a maximum length scale of $\approx$100\,nm.
The vicinal terrace structure is already discernible.
At 1100\,$^\circ$C during heating (Fig.~\ref{fig:thermalprep}(c)), the final terrace structure with a terrace width limited by the geometrical miscut of the sample has developed.
During the annealing at 1300\,$^\circ$C (d), terraces with uniform contrast in AFM arise.
The final morphology of the surface after cooling to the growth temperature of 840\,$^\circ$C is shown in Fig.~\ref{fig:thermalprep}(e).
If the miscut allows, large terraces of uniform width have formed, separated by almost straight steps.
Height profiles along the lines across the image (Fig.~\ref{fig:thermalprep}(g)) reveal a step height corresponding to the lattice constant of SrTiO$_3$, the low noise indicating a smooth and uniform terrace surface.
The phase contrast image in Fig.~\ref{fig:thermalprep}(f), acquired simultaneously with the image in (e), shows uniform intensity on the terraces, indicating a single, well-ordered surface reconstruction.
This result is virtually indistinguishable from SrTiO$_3$ substrate surfaces prepared with the traditional method (not shown).
Slowly cooling the substrate from the annealing to room temperature at 1\,K/s does not alter the terrace structure (Fig.~\ref{fig:thermalprep}(h)).

Reflection high-energy electron diffraction (RHEED) patterns acquired along $<$100$>$ during the heat treatment are shown in Fig.~\ref{fig:rheed}.
Panel (b) shows the pattern prior to heating, corresponding to the surface morphology of Fig.~\ref{fig:thermalprep}(a).
The pattern indicates a 2D surface, with sharp reflections on the Laue circle, weak intensity along the streaks and a clear Kikuchi line pattern.
The absence of fractional-order streaks due to surface reconstruction and the relatively low contrast (significant diffuse intensity between the streaks, low intensity of integer order reflections other than the specular spot, with the specular spot just reaching maximum intensity indicated by red color) are typical for small-scale disorder on the surface.

During heating, at the typical growth temperature of 850\,$^\circ$C (Fig.~\ref{fig:rheed}(c), corresponding to Fig.~\ref{fig:thermalprep}(b)), the surface mobility is sufficient to form atomically smooth islands and surface reconstructions.
The increased order manifests itself in narrower streaks and an increased overall contrast of the diffraction pattern.
On two nominally identical batches from the same vendor\cite{shinkosha}, we have observed two different reconstructions in this temperature range.
The first batch showed a weak 2$\times$ pattern with a soft onset at 500\,$^\circ$C, barely visible in Fig.~\ref{fig:rheed}(c), the second batch developed a clear 3$\times$ pattern at 825$\pm$5\,$^\circ$C.
For the first batch, the 2$\times$ surface reconstructions vanished above 1000\,$^\circ$C, for the second batch, the 3$\times$ surface reconstructions vanished above 1200\,$^\circ$C.
Both batches behaved identically from then on.
Within each batch, every sample showed the same reconstruction (2$\times$ for the first and 3$\times$ for the second) during heating (RHEED patterns in Fig.~\ref{fig:rheed}(c,d)).

Starting at 1100\,$^\circ$C, the reflections on the Laue circle begin to elongate.
Around 1200\,$^\circ$C, surface reconstructions vanish and the Kikuchi line pattern blurs.
The streaks remain thin and sharply defined.
This state is shown in Fig.~\ref{fig:rheed}(e), halfway through the 200\,s annealing plateau at 1300\,$^\circ$C.
During these 200\,s, a weak 2$\times$ reconstruction develops, which intensifies during cooling, and condenses into sharp reflections along the Laue circle below 1000\,$^\circ$C.

The final state at the deposition temperature of 860\,$^\circ$C is shown in Fig.~\ref{fig:rheed}(f).
This pattern, corresponding to the morphology of Fig.~\ref{fig:thermalprep}(e), is characterized by a high dynamic range (lower diffuse intensity than the previous patterns, with significantly overexposed specular spot), a crisp Kikuchi line pattern, and along the Laue circle sharp reflections that are often very close to circular.
As both the integer and the fractional order reflections are circular, both the terrace sizes and the reconstruction domain sizes on the terraces have reached or exceeded the resolution limit of RHEED.
The surface has attained a highly ordered state, ideal for subsequent epitaxy.
The RHEED pattern of a chemically prepared substrate is shown for reference in Fig.~\ref{fig:rheed}(g,h).
Both patterns were recorded at room temperature, the first one after introducing the substrate, the second one after an additional thermal preparation cycle to 1300\,$^\circ$C, after which it was slowly cooled down.
The pattern in (g) already shows intense integer order reflections along the Laue circle and a clear 2$\times$ reconstruction pattern, which is maintained throughout the thermal preparation cycle.
Slowly cooling from 860\,$^\circ$C usually leads to a more disordered reconstruction, which seems to become less stable at lower temperatures and often transforms to a 4$\times$ symmetry.

None of the thermally prepared substrates showed a color change after the annealing procedure ($p_{\mathrm{O}_2}$=0.075\,mbar), although identical samples heated to 1300\,$^\circ$C while maintaining the vacuum base pressure of the chamber were black.

To explore the suitability of thermally prepared SrTiO$_3$ (001) for epitaxy, we have grown both homoepitaxial and heteroepitaxial layers on these surfaces.
The example provided in Fig.~\ref{fig:growth} charcterizes the growth of a homoepitaxial layer grown at 860\,$^\circ$C and $p_{\mathrm{O}_2}$=7.5\,Pa (0.075\,mbar).
The laser repetition rate was 1\,Hz the energy density on the target 2\,J/cm$^2$.
The RHEED intensity oscillations (Fig.~\ref{fig:growth}(a)) decay during the first 20 oscillations to small oscillations around a constant average intensity, after which the growth proceeds with small further decay, both in amplitude and average intensity.
To count the oscillations, the intensity data were transformed and fitted\cite{braun2017real}.
The light blue curve plots the average intensity in the area indicated by the blue rectangle in the inserted RHEED image.
This curve was smoothed by a running average with a 1\,s window width corresponding to the laser repetition rate to yield the curve shown in dark blue.
The dark blue curve was then processed\cite{braun2017real} to remove the average intensity variation and the decay, resulting in the green plot.
A cosine fit (shown in black in Fig.~\ref{fig:growth}(a)) to the green curve indicates a total of 109 oscillations.
In the absence of step-flow contributions, this corresponds to a layer thickness of 109\,ML.
The AFM scan in Fig.~\ref{fig:growth}(b) reveals the surface morphology at the end of this growth, corresponding to the RHEED pattern shown in the inset of Fig.~\ref{fig:growth}(a).
The heating laser was turned off together with the last deposition pulse to suppress post-growth recovery.
Even after the deposition of more than 100\,ML, the originally prepared terrace structure is still visible, although the average feature size on the surface is now significantly smaller.
Anywhere on the surface, not more than three levels (two incomplete layers) are locally present, confirming the persistence of layer-by-layer growth for long growth times and the good quality of the substrate template for homoepitaxial growth.

Two series of heteroepitaxial LaAlO$_3$ layers of varying thickness (not shown) were deposited on SrTiO$_3$ (001) substrates prepared as described above.
In both series, no conductivity was observed for bare substrates and LaAlO$_3$ layer thicknesses below 3.5\,ML, whereas thicker layers were conductive with mobilities of 1470 and 1585\,cm$^2$/(Vs) and carrier concentrations of 2.0--2.3$\times$10$^{13}$\,cm$^{-2}$ for 4\,ML thick layers, all measured at 2\,K.

\begin{figure}
\includegraphics[width=\columnwidth]{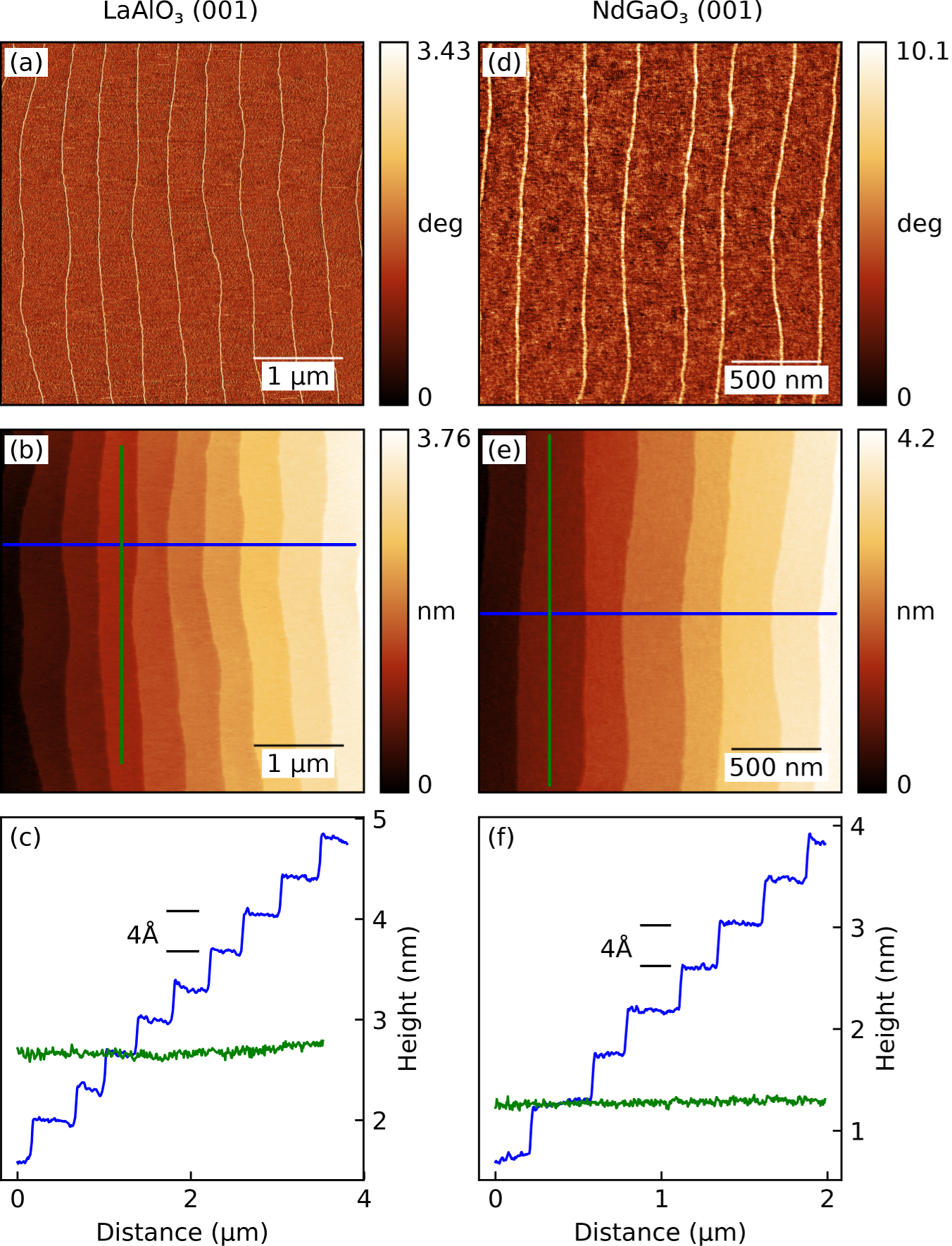}%
\caption{Thermally prepared substrate surfaces of (001)-oriented LaAlO$_3$ (left column, a--c) and (001)-oriented NdGaO$_3$ (right column, d--f).
                Topography is shown in (b) and (e), the phase signal (dissipation) is shown in (a) and (d).
                Blue and green lines in (b) and (e) indicate the locations where the profiles shown in (c) and (f) have been acquired.\label{fig:othersubstrates}}%
\end{figure}

We have applied the thermal preparation method also to LaAlO$_3$ (001) and NdGaO$_3$ (001) substrates.
The results are shown in Fig.~\ref{fig:othersubstrates}, again as AFM scans on substrates rapidly quenched from the preparation temperature by abruptly turning off the heating laser.

The left column in Fig.~\ref{fig:othersubstrates} presents the studies of LaAlO$_3$.
We have found that, as compared to SrTiO$_3$, a higher final temperature (1500\,$^\circ$C) and a lower oxygen background pressure (0.75\,Pa) yielded the best surfaces.
At lower temperatures, islands remained on the terraces.
At 7.5\,Pa oxygen pressure, the surface was found to be unstable when cooling it to T$\approx$1000\,$^\circ$C.
The surface topography (Fig.~\ref{fig:othersubstrates}(b)) shows terraces of very similar width, the terrace edges are less straight than on SrTiO$_3$.
Height profiles on the terraces and across the steps (Fig.~\ref{fig:othersubstrates}(c)) confirm smooth terraces with a monolayer step height consistent with the LaAlO$_3$ [001] pseudocubic lattice constant of 3.78\,\AA.
The phase contrast image acquired simultaneously with the topographic image (Fig.~\ref{fig:othersubstrates}(a)) shows uniform, virtually identical values on all terraces, confirming a uniform surface reconstruction.
RHEED (not shown) indicates a similar surface quality as on SrTiO$_3$, however no fractional-order streaks were observed.
Good homoepitaxy with persistent RHEED intensity oscillations similar to the results shown in Fig.~\ref{fig:growth}(a) was observed for growth at 1000\,$^\circ$C with 0.75\,Pa oxygen background pressure.

The results for NdGaO$_3$ are depicted in the right column of Fig.~\ref{fig:othersubstrates}.
For this material, a lower final temperature of 1000\,$^\circ$C at 7.5\,Pa oxygen pressure was found to yield the best surface morphology.
Again, the surface morphology (Fig.~\ref{fig:othersubstrates}(e)) shows terraces of very similar width as on SrTiO$_3$, this time with edges that are almost as straight as on SrTiO$_3$.
Height profiles on the terraces and across the steps (Fig.~\ref{fig:othersubstrates}(f)) again confirm smooth terraces with a monolayer step height consistent with the NdGaO$_3$ [001] pseudocubic lattice constant of 3.86\,\AA.
The phase contrast image acquired simultaneously with the topographic image (Fig.~\ref{fig:othersubstrates}(d)) displays a somewhat larger contrast variation, but still a fairly uniform intensity on the terraces.
RHEED (not shown) indicates a similar surface quality as on SrTiO$_3$, with 2$\times$ fractional-order streaks present throughout the entire preparation sequence.

Preliminary results on Al$_2$O$_3$ (0001) indicate that the thermal preparation method also works on this surface, with annealing temperatures $>$ 1500\,$^\circ$C.

In summary, we find that a SrTiO$_3$ (001) substrate preparation procedure consisting of degreasing steps, followed by a deionized water bath and an in-situ anneal at elevated temperatures in molecular oxygen at pressures of 1 to 10\,Pa, leads to surfaces that are at least equal in quality than the traditional ex-situ method with an HF etch and ambient pressure anneal.
This novel method is cleaner than the traditional one, since the main substrate preparation step takes place in the deposition chamber, directly before layer deposition.
It allows tighter parameter control, as it involves fewer parameters (oxygen partial pressure, substrate temperature) which may be entirely computer controlled.
The method is safer, as it avoids the use of HF or other aggressive and harmful chemicals.
It is also quicker, as the number and duration of the chemical cleaning steps is minimized, with the in-situ annealing procedure typically requiring half an hour or less.

The method has worked on many other oxide substrates we have studied so far.
It is therefore possible that it is applicable to a wider range of oxide substrates, potentially providing a general, standardized method that allows the straightforward use of a large variety of substrates in the same growth chamber with minimal external equipment.

The authors thank Hans Boschker, Prosper Ngabonziza and Pascal Wittlich for various contributions to this work.


%

\end{document}